\documentclass[journal]{IEEEtran}
\IEEEoverridecommandlockouts

\usepackage{graphicx}
 \usepackage{cite}
 \usepackage{array}
 \usepackage{amsmath, amssymb}
 \usepackage{lipsum}
\usepackage{multicol}
\usepackage{authblk}
\usepackage{lettrine}
\usepackage{float}
\usepackage{xcolor}

\usepackage{amsmath,amssymb}
\usepackage{tikz}
\usetikzlibrary{positioning,arrows.meta}

\newcommand{\change}[1]{\textcolor{black}{#1}}

\begin{document}

% \title{Anonymous Mutual Authentication with Flight Path Privacy Based on A Zero-Knowledge Proof for UAV Systems}
\title{ZAPS: A Zero-Knowledge Proof Protocol for Secure UAV Authentication with Flight Path Privacy}

\author[1,3]{Shayesta Naziri}
\author[1]{Xu Wang}
\author[1]{Guangsheng Yu}
% \author[2]{Jian Xu}
\author[2]{Christy Jie Liang}
\author[1]{Wei Ni}

\affil[1]{School of Electrical and Data Engineering, University of Technology Sydney, NSW, 2007, AUS}
\affil[2]{School of Computer Science, University of Technology Sydney, NSW, 2007, AUS}
\affil[3]{Corresponding Author: Shayesta Naziri ( Shayesta.Naziri@student.uts.edu.au)}

\setlength{\columnsep}{0.24in}
\def\BibTeX{{\rm B\kern-.05em{\sc i\kern-.025em b}\kern-.08em
    T\kern-.1667em\lower.7ex\hbox{E}\kern-.125emX}}

\markboth{Journal of \LaTeX\ Class Files,~Vol.~14, No.~8, Jun~2025}%
{Shell \MakeLowercase{\textit{et al.}}: Bare Demo of IEEEtran.cls for IEEE Journals}

\maketitle

\begin{abstract}
 The increasing deployment of Unmanned Aerial Vehicles (UAVs) for military, commercial, and logistics applications has raised significant concerns regarding flight path privacy. Conventional UAV communication systems often expose flight path data to third parties, making them vulnerable to tracking, surveillance, and location inference attacks. Existing encryption techniques provide security but fail to ensure complete privacy, as adversaries can still infer movement patterns through metadata analysis. To address these challenges, we propose a zk-SNARK (Zero-Knowledge Succinct Non-Interactive Argument of Knowledge)-based privacy preserving flight path authentication and verification framework. Our approach ensures that a UAV can prove its authorisation, validate its flight path with a control centre, and comply with regulatory constraints without revealing any sensitive trajectory information. By leveraging zk-SNARKs, the UAV can generate cryptographic proofs that verify compliance with predefined flight policies while keeping the exact path and location undisclosed. This method mitigates risks associated with real-time tracking, identity exposure, and unauthorised interception, thereby enhancing UAV operational security in adversarial environments. Our proposed solution balances privacy, security, and computational efficiency, making it suitable for resource-constrained UAVs in both civilian and military applications.

\end{abstract}


\IEEEpeerreviewmaketitle

\section{Introduction}
\lettrine{U}{nmanned} Aerial Vehicles (UAVs) are autonomous flying robots that vary in size and capability, from long-range UAVs that can cover hundreds of miles to small, agile drones built for operating in tight spaces~\cite{ref1}.
The rapid advancement of drone technologies has paved the way for a new generation of intelligent, autonomous delivery systems.
These systems are increasingly being adopted across a wide spectrum of domains including e-commerce, healthcare, agriculture, and defense for their ability to provide fast, efficient, and contactless delivery. As UAV technology becomes more embedded in urban and sensitive operational environments, the need to secure drone communications and preserve user and mission privacy is growing more critical than ever~\cite{ref2,ref3}.

Among the most pressing concerns in secure drone delivery systems is the flight path privacy problem. Unlike conventional delivery mechanisms, UAVs operate in open-air spaces and are guided by flight plans that often include real time updates, geolocation tracking, and route-specific telemetry. The disclosure of such information whether through intercepted messages, network leaks, or internal compromise can expose sensitive details such as the origin and destination of deliveries, user identities, operational timings, and overall mission intent. This not only endangers user privacy but also invites a range of attacks such as drone hijacking, delivery interception, location profiling, and traffic analysis~\cite{ref4,ref5}. 

In drone delivery ecosystems, multiple entities typically participate in mission coordination, including the user (sender/receiver), the drone, and a central server or ground station. In traditional designs, these entities exchange cryptographic messages for authentication and coordination, but their identities and the drone’s trajectory are often either shared or inferable~\cite{ref6}. Even if data encryption is applied, adversaries can perform linkability attacks, timing analysis, or traffic fingerprinting to correlate entities and deduce flight paths. Additionally, adversaries with access to the server or drone can passively harvest metadata to reconstruct the delivery pattern and compromise both identity and location privacy. Insider threats, where  entities like the server seek to extract private information, require the protocol to protect privacy even among participants~\cite{ref7,ref8}. This sets a high bar for privacy-preserving design, where even legitimate participants must be unable to link identities to routes or extract location details.

Existing solutions to drone security predominantly focus on securing communication channels using symmetric or public-key encryption, message authentication codes, or lightweight key exchange protocols such as ECDH~\cite{ref9}. However, these methods primarily ensure confidentiality and authenticity without providing guarantees of nondisclosure of the underlying route or unlinkability between communicating parties. Furthermore, most traditional protocols lack mechanisms for verifiable computations without revealing sensitive data, making them unsuitable for adversarial environments where zero-trust principles are required.

To address these limitations, we design a protocol that authenticates participants and secures communication while allowing them to prove correct behavior without revealing private data, such as user identities or the drone’s flight path. This motivates the use of zero-knowledge proofs specifically, zk-SNARKs which enable proving knowledge of a statement without exposing the underlying data \cite{ref11}.
We propose a privacy-preserving drone delivery protocol that integrates Elliptic Curve Diffie-Hellman (ECDH) for efficient key exchange and zk-SNARKs for identity and route privacy. The protocol ensures mutual authentication among the User, Drone, and Server, while preserving critical privacy and security guarantees.
Flight Path Privacy: The drone's route is not exposed in plaintext; instead, zk-SNARKs are used to prove route correctness without disclosure.

\begin{enumerate}

\item The proposed protocol achieves robust flight path privacy through the integration of cryptographic commitments and zero-knowledge proofs, specifically zk-SNARKs. These cryptographic tools allow the drone to prove the correctness of its route without disclosing any portion of the path in plaintext. To further enhance unlinkability, each communication session is initialized with fresh, randomized parameters, including ephemeral keys and unique proofs, ensuring that interactions remain independent and resistant to correlation. 

\item A key strength of the proposed protocol lies in its robust privacy preserving architecture, which is specifically designed to ensure that the identities of all involved entities, namely the user, drone, and server, remain completely hidden throughout the communication process. This anonymity is maintained through a combination of cryptographic techniques: secure communication channels are established using Elliptic Curve Diffie-Hellman (ECDH) based key exchange, followed by the application of symmetric encryption to protect data integrity and confidentiality. Additionally, such as the drone’s flight path, the protocol leverages zero-knowledge proofs (zk-SNARKs), enabling the verification of route correctness without revealing the actual path and identity protection, making the protocol highly suitable for scenarios demanding stringent privacy guarantees. 
\end{enumerate}

The remainder of this paper is organized as follows: Section~\ref{sec_relatedwork} presents related work in drone authentication and privacy-preserving communication. Section \text{III} outlines the system model, design and goals. Section \text{IV} details the construction of our proposed protocol, including the ECDH key agreement and zk-SNARK circuit generation. Section \text{V} evaluates the protocol’s security, privacy, and performance. Finally, Section \text{VI} concludes with future directions.

\section{Related Work}
\label{sec_relatedwork}
Flight path privacy, like location privacy in traditional mobile networks, is a crucial concern in the development of UAV systems\cite{ref10}. However, as there is a lack of protocols specifically designed to protect flight path privacy. To fill this void, we will assess existing protocols that focus on location and other privacy aspects to understand how they might be adapted or extended to ensure better protection of drone flight paths.  
Various authentication protocols have been proposed to bolster UAV security. For example, \cite{ref12} combines non-interactive zero-knowledge proofs (NIZKP) with bilinear mapping to achieve unlinkable user identities and resistance to malleability attacks, while \cite{ref13} employs distance-bounding protocols and anonymous certificates to verify UAV locations without compromising privacy. The multi-factor authentication schemes proposed in \cite{ref14} and \cite{ref15} integrate components such as passwords, biometrics, and smart cards to strengthen identity verification. However, they fall short in addressing essential requirements, including replay attack resistance, credential revocation, and protection of location privacy. The authors in \cite{ref16} implemented a registration hub-based identity management system within the Industrial IoT context, which, despite offering structured identity handling, does not mitigate location privacy concerns and introduces additional overhead. Similarly, the authors of work \cite{ref17} employed mobile edge computing with online-offline signatures in their lightweight UAV authentication protocol, but their design still experiences limitations due to constrained memory and processing power.

Moving beyond traditional encryption schemes, in the proposed protocol \cite{ref18} applied AI techniques to drone surveillance, allowing for feature recognition and tracking over wireless channels. Still, their model experiences significant latency and does not offer strong security guarantees. Similarly, work in \cite{ref19} and \cite{ref20} focused on optimizing drone delivery and route planning, but often at the expense of scalability and privacy protections. To address certificate overhead, in the protocol \cite{ref21} proposed a certificateless approach adaptable to multiple communication settings, including conditional tracking. 
Nonetheless, it does not adequately secure location privacy and results in high communication costs in batch operations. The works of authors \cite{ref22} introduces a lightweight security protocol designed to ensure tamper-resistant data exchanges between UAVs and ground control stations, emphasizing integrity and confidentiality. 
 In \cite{ref23} proposes a secure communication framework to mitigate threats posed by malicious UAVs, ensuring consensus reliability. The work in~\cite{ref24} employs automated algorithms for mission-specific swarm formation selection, streamlining coordination in dynamic environments. In \cite{ref25}, developed a mutual authentication protocol enabling direct communication between drones and users, eliminating the need for intermediaries. However, their scheme is still prone to impersonation and man in the-middle attacks. The authors in~\cite{ref26} put forward a smart card-based authentication method, which, while reasonably secure, is vulnerable to password guessing and physical tampering.  In the study~\cite{ref27}, they employed elliptic curve cryptography (ECC) to enhance security, but their design involves considerable computational and storage demands due to complex key handling.
 
Efforts toward lightweight security mechanisms have also been notable. The authors in \cite{ref28} and \cite{ref29} proposed using physically unclonable functions and hash-based key agreements to counter standard attacks. Nevertheless, both fail to address the location privacy issue. ECC-based and multi-factor authentication protocols by the authors of works \cite{ref30} and \cite{ref31} offer resilience against multiple threats but introduce performance trade-offs and do not tackle location-based vulnerabilities.

In summary, while many current solutions excel in either security enhancement or performance optimization, few successfully achieve both while ensuring drone flight privacy. This work addresses this shortcoming by presenting a secure, efficient, and flight-privacy-preserving authentication protocol designed IoD applications.

\section{Preliminaries}
\subsection{Zero Knowledge Proof}\label{AA}
A zero-knowledge proof (ZKP) is a cryptographic protocol where a prover confirms the truth of a statement to a verifier while ensuring that no other information is leaked\cite{ref32}. Zero-knowledge proofs fall into two main categories: interactive and non-interactive. In interactive proofs, the prover and verifier communicate multiple times, whereas in non-interactive proofs, the prover provides a single proof that the verifier can check on its own. If a drone uses an interactive proof with an edge server, it will have to hover longer due to the repeated communication steps, leading to greater energy consumption~\cite{ref33}.

 \subsubsection{Zk-SNARKs} A zero-knowledge succinct non-interactive ar
arguments of knowledge (zk-SNARKs) scheme represents a specific form of Zero-Knowledge Proof (ZKP) that enables a prover to convince a verifier of their knowledge of a witness, without disclosing any details about the witness itself. This scheme is defined by four key algorithms: Setup, Keygen, Proof, and Verify~\cite{ref34}.

 Setup$(1^\lambda) \to \mathcal{Z}$: This algorithm takes as input a security parameter $\lambda$ and gives as output a set of public parameters $\mathcal{Z} = \{e, p, g_1, g_2, \mathbb{G}_1, \mathbb{G}_2, \mathbb{G}_T\}$, where $p$ is a prime number, $e: \mathbb{G}_1 \times \mathbb{G}_2 \to \mathbb{G}_T$ is a bilinear map ~\cite{ref35}, and where $\mathbb{G}_1$ and $\mathbb{G}_2$ are acyclic groups (p-order) with generators $g_1$ and $g_2$, respectively.

Keygen$(C) \to (Pk, Vk)$: The Keygen algorithm takes as an input the arithmetic circuit and uses the public parameters $\mathcal{Z}$ to generate a pair of keys for proving and verifying the statement: $(Pk, Vk)$.

Genproof$(Pk, x, w) \to \pi$: The Genproof algorithm takes as input the proof key that was generated by the Keygen algorithm, the statement $x$ (input of circuit $C$) and a secret witness $w$ (auxiliary input of circuit $C$), and generates a zero-knowledge proof $\pi$ based on the relation between the circuit $C$, the statement $x$ and the witness $w$.

Verproof$(Vk, x, \pi) \to b$: The Verproof algorithm takes as input the verification key $Vk$, the statement $x$, and the proof $\pi$ and generates as output a binary number based on the proof’s $\pi$ validity. If the proof is valid, then $b=1$ or else $b=0$.

The zk-SNARK scheme satisfies the following properties:

Completeness: If a statement is $x \in \mathcal{L}$, and $w$ is a valid witness of $x$, then the verifier accepts the proof with probability 1.
\[
\Pr[\text{Verproof}(Vk, x, \pi) \to 1 \mid \text{Genproof}(Pk, x, w) \to \pi] = 1
\]

Zero knowledge: Without revealing any information regarding the witness $w$, the prover can prove to the verifier that the statement $x$ is true. This can be described mathematically as follows: Let $\mathcal{S}$ be a simulator that, given a statement, $x \in \mathcal{L}$ and the $Pk$ can produce a proof that is indistinguishable from a real proof generated by the prover, without knowing the witness $w$. Then,
\[
\{\text{Genproof}(Pk, x, w)\} \approx \{\mathcal{S}(Pk, x)\},
\]
where “$\approx$” denotes computational indistinguishability.

Soundness: If a witness $w$ is not valid, then a malicious actor cannot craft a proper proof. If we denote the generated malicious proof with $\tilde{\pi}$, then
\[
\Pr[\text{Verproof}(Vk, x, \tilde{\pi}) \to 1] \leq \epsilon,
\]
where $\epsilon$ negligible.

\subsubsection{Hash Function} We use a hash function 
\[
H: \{0,1\}^* \rightarrow \{0,1\}^\lambda
\]
where \( \lambda \) is a positive integer determined by the security parameter. This function is assumed to have the following key property:

{Collision Resistance:} It should be computationally infeasible for an adversary to find two distinct inputs \( m_1 \) and \( m_2 \) such that
\[
H(m_1) = H(m_2).
\]

Furthermore, we rely on the fact that if a hash function is collision-resistant, it also provides the related security properties of \textit{one-wayness} (hard to invert) and \textit{second preimage resistance} (hard to find a different input that hashes to the same output).

\subsubsection {Elliptic Curve Diffie-Hellman (ECDH)}
Given an elliptic curve \( E \) over a finite field \( \mathbb{F}_q \), and a base point \( G \in E(\mathbb{F}_q) \) of prime order \( n \), two parties A and B wishing to share a secret key \( K \) do so as follows:

\begin{itemize}
    \item A selects a random scalar \( a \in \mathbb{Z}_n \) and sends \( M_A = aG \) to B.
    \item B selects a random scalar \( b \in \mathbb{Z}_n \) and sends \( M_B = bG \) to A.
    \item A computes \( K = aM_B = abG \), and B computes \( K = bM_A = abG \).
\end{itemize}

Both parties arrive at the same shared secret \( K = abG \), which can then be used as a symmetric key or further processed using a Key Derivation Function (KDF).   

\section{System and Proposed Protocol}
\label{sec_system}
\subsection{Main Entities}
\label{sec_entities} 
\begin{figure}[h]
    \centering
    \includegraphics[width=0.99\linewidth]{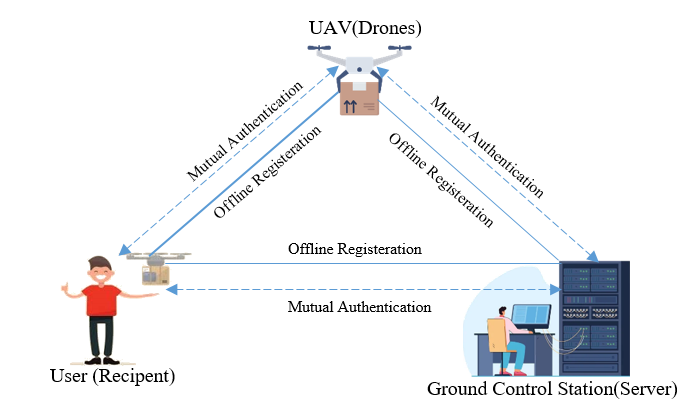}
    \caption{System Model of the ZAPS protocol.}
    \label{fig_entities}
\end{figure}

As illustrated in Fig.~\ref{fig_entities}. in the proposed drone delivery system, there are three main entities: the User (U), the Drone (D), and the Server (S). Their interactions and roles are depicted in the protocol described above~\cite{ref36,ref37}.

\begin{itemize}
    \item User (U): The User/Recipient in a drone delivery system is the central entity that initiates the delivery request, communicates securely with the Server and Drone, and confirms the receipt of the delivered item. 
    
    \item Drone (D): The Drone in a drone delivery system is a critical component responsible for physically transporting goods, communicating with the Server and User, and ensuring the safe and secure delivery of packages. Its role is essential in facilitating the entire delivery process, from initialization to final acknowledgment.
    
    \item Server (S): In a drone delivery system, the Server can also be considered as a Ground Control Station or Sender, which plays a crucial role in managing and coordinating the entire delivery process. This server acts as the central hub for communication, control, and data management. It is responsible for generating and managing cryptographic keys, verifying authentication tokens, and ensuring secure communication between the User, the Drone, and itself. The Ground Station Server calculates optimal flight paths for the Drone, monitors its status in real-time, and adjusts flight plans as needed to address any unexpected situations. Additionally, it stores and manages all relevant data, including user information, drone status, and delivery requests, ensuring efficient order processing and transaction management.
\end{itemize}

\subsection{System Description}
\label{sec_sysdes}
The proposed secure drone delivery system consists of three main entities: the \text{Server}, the \text{Drone}, and the \text{Recipient (User)}, working together to ensure privacy-preserving and authenticated delivery using elliptic curve cryptography (ECDH) and zk-SNARKs. The zk-SNARK parameters---specifically the proving key $P_k$ and verification key $V_k$---are generated during a one-time trusted setup phase conducted entirely offline prior to system deployment. This trusted setup is assumed to be carried out in a secure environment, and any toxic waste (intermediate randomness) is securely discarded to prevent compromise. Once generated, $P_k$ and $V_k$ are securely distributed to the corresponding system entities.

As illustrated in Fig.~2, the process begins when a delivery is initiated, and the Server initializes both the Drone and Recipient by registering their identities and cryptographic credentials. Once initialization is successful, secure keys are exchanged between the Drone and the Recipient using an ECDH-based key agreement to establish a shared session key for encrypted communication.

To ensure privacy of sensitive delivery details, such as the Recipient's identity and the Drone's flight path, the Recipient generates a zk-SNARK proof $(\pi)$ that validates delivery authorization without revealing any private data. This proof is then verified by the Drone or Server using the public verification key $V_k$. If the zk-SNARK proof is valid, the Drone proceeds with the delivery. Finally, the Recipient confirms the secure reception of the package using a cryptographically signed acknowledgment.

This end-to-end protocol ensures strong authentication, delivery integrity, anonymity, and flight path confidentiality while operating under clearly defined trusted setup assumptions.

\begin{figure}[h]
    \centering
    \includegraphics[width=1\linewidth]{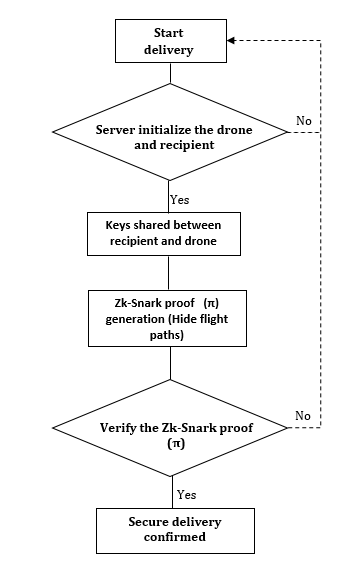}
    \caption{Flowchart of the ZAPS protocol}
    \label{fig_flowchart}
\end{figure}

\begin{table}[!htbp]
  \centering
  \caption{Notation used in this protocol}
  \label{tab_notations}
  \begin{tabular}{c|l}
    \hline
    \textbf{Notation} & \textbf{Description} \\ \hline
    $U_{u}$                 & \textit{$u$ User (Recipient)} \\ 
    $GCS_{s}^{\,s}$       & \textit{$s^{\text{th}}$ Ground Control Station (Server)} \\ 
    $D_{d}$                 & \textit{$d^{\text{th}}$ Drone} \\ 
    $ID_{i},\,PW_{i}$       & \textit{Identity, Password} \\ 
    $\pi$                   & \textit{zk‑SNARK proof} \\ 
    $C_{i}$                 & \textit{Commitment value} \\ 
    $W_{i}$                 & \textit{Private witness (e.g.\ flight path)} \\ 
    $r_{i}$                 & \textit{Random value} \\ 
    $V_{k,i},\,P_{k,i}$       & \textit{Verification and proving keys for proofs} \\ 
    $V,\,P$                 & \textit{Private and Public key} \\ 
    $G$                     & \textit{Base point of elliptic curve} \\
    $\mathbb{Z}_{p}^{\ast}$ & \textit{Multiplicative group of order $p-1$} \\ 
    $\mathbb{F}_{p}$        & \textit{Finite field} \\ 
    $E_{p}(a,b)$            & \textit{Elliptic curve over $\mathbb{F}_{p}$} \\ 
    $\Delta T$              & \textit{Maximum transmission delay} \\ 
    $CT$                    & \textit{Current timestamp} \\ 
    $\mathrm{Auth}$         & \textit{Verification information} \\ 
    $I$                     & \textit{Hash digest} \\ 
    $SK$                    & \textit{Session key} \\ 
    $SMK$                   & \textit{Symmetric key} \\ 
    $\tilde{A}$             & \textit{The adversary} \\ 
    \textit{ECDH}           & \textit{Elliptic‑Curve Diffie–Hellman} \\ 
    $h(\cdot)$              & \textit{One‑way hash function} \\ 
    $\parallel$             & \textit{Concatenation} \\ \hline
  \end{tabular}
\end{table}

\subsection{Proposed Protocol}
\label{sec_protocol}
The protocol employs a four-phase to ensure secure and private operations in drone delivery systems: (1) System Setup Phase, (2) Registration Phase, (3) Initialization and authentication and (4) zk-SNARK proofs generation for flight path privacy. Furthermore, the notations with their significance tabulated in Table~\ref{tab_notations} are utilized in describing as well as analyzing protocol contains the following phases.

\subsubsection {Setup Phase} In this phase, the Control Authority (CA) initializes system-wide parameters and cryptographic elements. The server or Ground Control Station (GCS) begins by selecting a non-singular elliptic curve \( E_p(a, b) \) defined over a prime finite field \( \mathbb{Z}_p \), where \( p \) is a large prime satisfying the condition \( 4a^3 + 27b^2 \not\equiv 0 \pmod{p} \). A base point \( G \) on the curve is chosen with a prime order \( n \), such that \( n \cdot G = \mathcal{O} \), where \( \mathcal{O} \) denotes the point at infinity. The server selects a private key \( V_s \in \mathbb{Z}_p \) and computes its corresponding public key \( P_s = V_s \cdot G \). A collision-resistant cryptographic hash function \( H(\cdot) \), such as SHA-256, is also chosen. The public system parameters \( \{ E_p(a, b), p, G, H(\cdot), P_s \} \) are published, while the private key \( V_s \) remains confidential and is securely stored by the server.

\subsubsection {Registration Phase} All participating entities (e.g., User, Drone, and Server) are securely registered by the CA before network deployment. The Control Authority securely registers users and drones with the server in an offline environment. For each user \( U_i \), a unique identity \( ID_u \) is assigned. 

A private key \( V_u \in \mathbb{Z}_p^* \) is generated, and the corresponding public key is computed as \( P_u = V_u \cdot G \). A symmetric key \( SMK_{u,s} = V_u \cdot P_s \) is derived for secure communication with the server. The information set \( \{ ID_u, V_u, SMK_{u,s}, H(\cdot) \} \) is securely pre-loaded into the user's device. Similarly, for each drone \( D_i \), a unique identity \( ID_d \) is assigned, and a private key \( V_d \in \mathbb{Z}_p^* \) is selected. The corresponding public key is computed as \( P_d = V_d \cdot G \). The server \( S_i \) independently generates its own private key \( V_s \in \mathbb{Z}_p^* \), computes the public key \( P_s = V_s \cdot G \), and publishes \( P_s \) as part of the public parameters.

\begin{figure*}[h] % 't' = top of the page
  \centering
  \includegraphics[width=\textwidth]{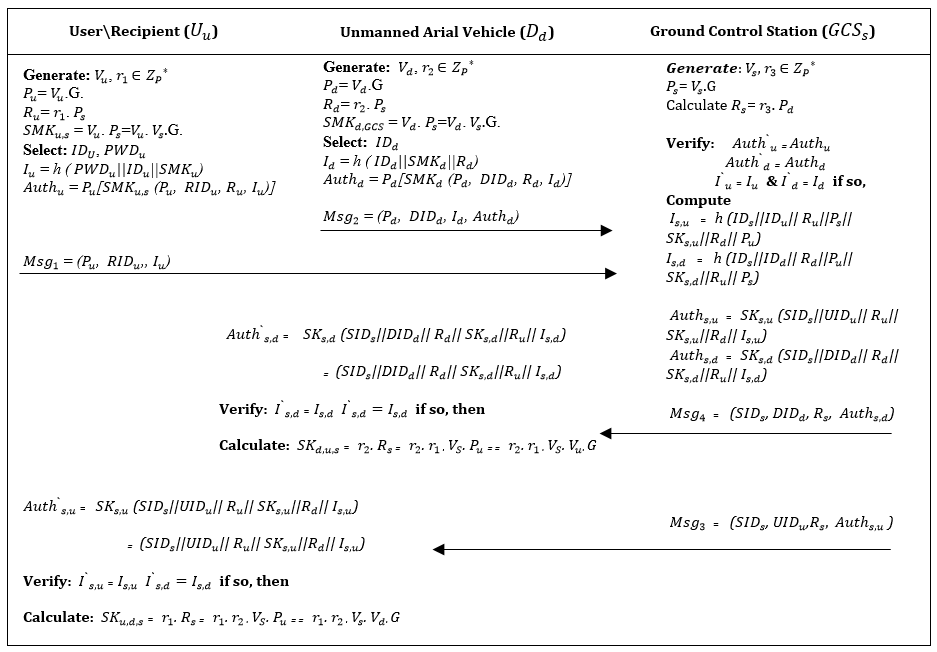}
  \caption{Initialisation and authentication phase of the ZAPS protocol.}
  \label{fig:wide_imageI}
\end{figure*}

\begin{figure*}[h] % 't' = top of the page
  \centering
  \includegraphics[width=\textwidth]{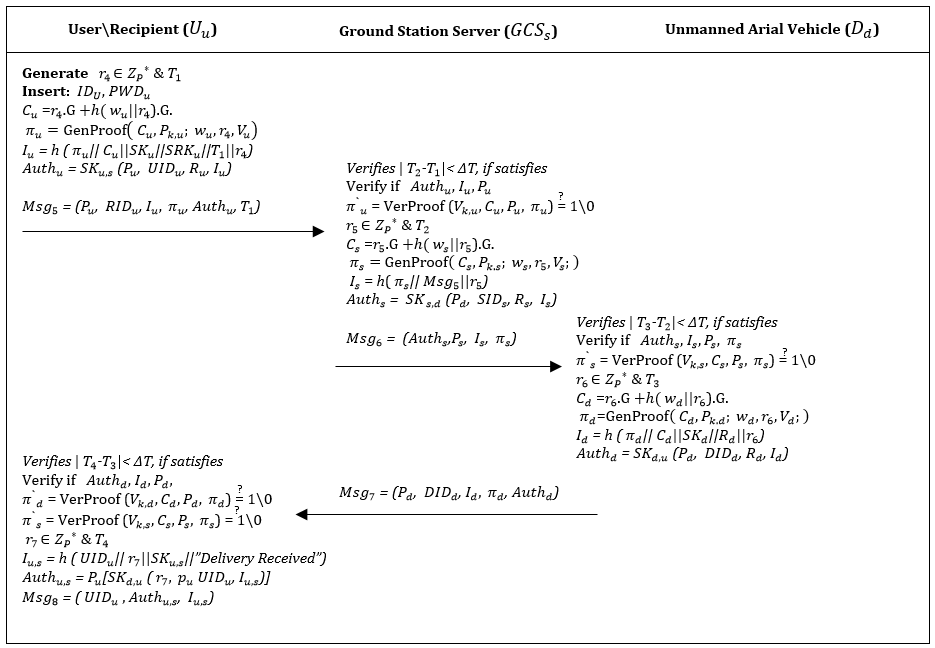}
  \caption{Zk-SNARK Proofs Generation Phase for Flight Path
Privacy for the ZAPS protocol.}
  \label{fig:wide_image}
\end{figure*}

\subsubsection {Initialisation and Authentication Phase} The initialization phase of our proposed protocol follows a similar approach to that described in \cite{ref38}, enhancing it to meet the unique security and privacy demands of our protocol.

In this phase, all registered entities mutually authenticate each other using ECDH-based exchange necessary cryptographic elements to enable zk-SNARK-based privacy-preserving operations. The user begins by selecting a random nonce \( r_1 \in \mathbb{Z}_p^* \) and a secret \( V_u \). The user computes their public key \( P_u = V_u \cdot G \), an ephemeral value \( R_u = r_1 \cdot P_s \), and the symmetric key \( \mathsf{SMK}_{u,s} = V_u \cdot V_s \cdot G \). Using these values, the user calculates a hashed identity commitment \( I_u = H(\mathsf{PWD}_u \, \| \, \mathsf{ID}_u \, \| \, \mathsf{SMK}_{u,s}) \), and constructs the authentication token \( \mathsf{Auth}_u = P_u \, \| \, \mathsf{SMK}_{u,s}(P_u, \mathsf{RID}_u, R_u, I_u) \). The user then sends message \( \mathsf{Msg}_1 = \{P_u, \mathsf{RID}_u, I_u\} \) to the server. 

Meanwhile, the drone chooses a random nonce \( r_2 \in \mathbb{Z}_p^* \) and a secret value \( V_d \). It computes its public key \( P_d = V_d \cdot G \), the ephemeral value \( R_d = r_2 \cdot P_s \), and the symmetric key \( \mathsf{SMK}_{d,s} = V_d \cdot V_s \cdot G \). The drone then computes \( I_d = H(\mathsf{ID}_d \, \| \, \mathsf{SMK}_{d,s} \, \| \, R_d) \), and forms its authentication message \( \mathsf{Auth}_d = P_d \, \| \, \mathsf{SMK}_d(P_d, \mathsf{DID}_d, R_d, I_d) \). The drone sends \( \mathsf{Msg}_2 = \{P_d, \mathsf{DID}_d, I_d, \mathsf{Auth}_d\} \) to the server.

Upon receiving both messages, the server proceeds to verify \( \mathsf{Auth}_u \) and \( \mathsf{Auth}_d \), then computes the authentication hashes \( I_{s,u} = H(\mathsf{ID}_s \, \| \, \mathsf{ID}_u \, \| \, R_u \, \| \, P_s \, \| \, \mathsf{SK}_{s,u} \, \| \, R_d \, \| \, P_u) \) and \( I_{s,d} = H(\mathsf{ID}_s \, \| \, \mathsf{ID}_d \, \| \, R_d \, \| \, P_u \, \| \, \mathsf{SK}_{s,d} \, \| \, R_u \, \| \, P_s) \), generating response tokens \( \mathsf{Auth}_{s,u} \) and \( \mathsf{Auth}_{s,d} \). The server then replies with \( \mathsf{Msg}_3 = \{\mathsf{SID}_s, \mathsf{UID}_u, R_s, \mathsf{Auth}_{s,u}\} \) to the user and \( \mathsf{Msg}_4 = \{\mathsf{SID}_s, \mathsf{DID}_d, R_s, \mathsf{Auth}_{s,d}\} \) to the drone.

In the final key derivation step, the drone calculates the session key as \( \mathsf{SK}_{d,u,s} = r_2 \cdot r_1 \cdot V_s \cdot V_u \cdot G \), while the user computes \( \mathsf{SK}_{u,d,s} = r_1 \cdot r_2 \cdot V_s \cdot V_d \cdot G \). Since the computations are equivalent, both parties derive the same mutual session key for secure communication. The specific steps are illustrated in Fig.~\ref{fig:wide_imageI}.

\subsubsection {zk-SNARK Proofs Generation Phase for Flight Path Privacy} Once mutual authentication is complete, this phase enables entities (e.g., Drone) to prove knowledge of sensitive flight path data without revealing the actual path using zk-SNARKs. As steps are illustrated
in Fig.~\ref{fig:wide_image}, the user generates a random nonce $r_4 \in \mathbb{Z}_P^*$, a timestamp $T_1$, and constructs a zk-SNARK circuit to encode their private flight path $w_u$ (treated as a secret witness). Using a Pedersen commitment scheme, the user computes $C_u = r_4 \cdot G + H(w_u \parallel r_4) \cdot G$, where $G$ is the elliptic curve base point. Leveraging tools like Circom and SnarkJS, the user generates a zero-knowledge proof $\pi_u = \text{GenProof}(C_u, P_{k,u}; w_u, r_4, V_u)$ based on precomputed trusted setup parameters. The proof $\pi_u$, along with public inputs, is authenticated via $Auth_u = SK_{u,s} \cdot (P_u, UID_u, R_u, I_u)$, where $I_u = H(\pi_u \parallel C_u \parallel SK_u \parallel SRK_u \parallel T_1 \parallel r_4)$. The user then dispatches $Msg_5 = (P_u, RID_u, I_u, \pi_u, Auth_u, T_1)$ to the server. 

The \textit{Server Action Phase} begins by verifying the freshness of $T_1$ using $|T_2 - T_1| < \Delta T$ to thwart replay attacks. The server validates $\pi_u$ via $\text{VerProof}(V_{k,u}, C_u, P_u, \pi_u)$, generates its own nonce $r_5 \in \mathbb{Z}_P^*$, and computes a commitment $C_s$ with proof $\pi_s$. It derives $I_s = H(\pi_s \parallel Msg_5 \parallel r_5)$ and signs $Auth_s = SK_{s,d} \cdot (P_d, SID_s, R_s, I_s)$, transmitting $Msg_6 = (Auth_s, P_s, I_s, \pi_s)$ to the drone.

In the \textit{Drone Action Phase}, the drone verifies the timestamp $T_2$ against $|T_3 - T_2| < \Delta T$, checks the server’s proof $\pi_s$, and generates a nonce $r_6 \in \mathbb{Z}_P^*$. It computes its commitment $C_d$, proof $\pi_d$, and authentication token $Auth_d = SK_{d,u} \cdot (P_d, DID_d, R_d, I_d)$, where $I_d = H(\pi_d \parallel C_d \parallel SK_d \parallel R_d \parallel r_6)$. The drone sends $Msg_7 = (P_d, DID_d, I_d, \pi_d, Auth_d)$ to the user.

Finally, during \textit{User Confirmation}, the user verifies $|T_4 - T_3| < \Delta T$, validates $\pi_d$ and $\pi_s$, and confirms delivery by generating an acknowledgment token $Auth_{u,s}$ to complete the protocol.

\section{Analysis}

We present a detailed examination of the privacy and security properties of the proposed scheme in this section. Additionally, we have conducted a thorough examination of the correctness and other security-related concerns of the proposed scheme.

\subsection*{A. Correctness Analysis}

\subsubsection*{1) Correctness of Initialization Phase}

In this phase, a session key correctly derived by the User $U$, \textit{Drone} $D$, \textit{and Server} $S$ \textit{using ECDH. Let} $G$ be the generator of the elliptic curve group $G$, and let $V_u, V_d, V_s \in \mathbb{Z}_p^*$ be the long-term private keys of the User, Drone, and Server respectively. Their corresponding public keys are defined as $P_u = V_u \cdot G$, $P_d = V_d \cdot G$, and $P_s = V_s \cdot G$. Let $r_1, r_2, r_3 \in \mathbb{Z}_p^*$ be fresh nonces used during the ECDH protocol. The ephemeral key from the User to the Server is $R_1 = r_1 \cdot P_s = r_1 V_s G$, from the Drone to the Server is $R_2 = r_2 \cdot P_s = r_2 V_s G$, and from the Server to the Drone is $R_s = r_3 \cdot P_d = r_3 V_d G$. The session key at the User side is computed as $SK_u = r_1 \cdot R_2 = r_1 \cdot (r_2 V_s G) = r_1 r_2 V_s G$, and at the Drone side as $SK_d = r_2 \cdot R_1 = r_2 \cdot (r_1 V_s G) = r_2 r_1 V_s G = r_1 r_2 V_s G$. Similarly, the session key at the Server side (using both nonces) is $SK_s = V_s \cdot r_1 r_2 G = r_1 r_2 V_s G$. Therefore, $SK_u = SK_d = SK_s = r_1 r_2 V_s G$. The correctness of the Initialization Phase has been proven.\\

\subsubsection*{2) Correctness of zk-SNARK Proof Generation Phase}

This protocol uses zk-SNARKs to prove a secret such as a flight path is valid without revealing it. We will break down the process step by step, covering the logic, math, and meaning behind the proof's correctness.
In this phase, the goal is to show that the zk-SNARK proof verifies that the committed values correspond to valid private data without revealing it. Let:

\begin{itemize}
    \item $w_i$: private witness (e.g., secret flight path, credentials)
    \item $r_i$: random nonce for hiding the witness
    \item $C_i = r_i \cdot G + H(w_i \parallel r_i) \cdot G$: Pedersen-style commitment
    \item $\pi_i = \text{Prove}(C_i, w_i, r_i, P_i)$: zk-SNARK proof generated by prover $i$
    \item $V_{k,i}$: verification key for $\pi_i$
\end{itemize}

 The verifier checks if the zk-SNARK proof is valid using a pairing based equation:
\begin{equation}
    \text{Verify}(vk_i, C_i, \pi_i) \Rightarrow e(VK + \pi_A, \pi_B) = e(\pi_H, vk_z) \cdot e(\pi_C, h)
\end{equation}
which ensures three critical things simultaneously: that the proof $\pi$ was generated correctly, that the values committed in $C_i$ are consistent with the private witness $w_i$, and that all computations align with the structure defined during the trusted setup phase.

The verification key $VK$ is constructed as
\begin{equation}
    VK = A_0(\tau) \cdot \rho_A \cdot G + \sum_{i=1}^{n} s_i A_i(\tau) \cdot \rho_A \cdot G
\end{equation}
where it aggregates polynomial evaluations at a secret toxic point $\tau$ scaled by $\rho_A$. The proof components $\pi_A, \pi_B, \pi_C$, and $\pi_H$ satisfy the equation
\begin{equation}
    \begin{split}
        e\big(A(\tau) \cdot \rho_A \cdot G,\ &B(\tau) \cdot \rho_B \cdot h\big) \\
        &= e(\pi_H, vk_z) \cdot e(\pi_C, h)
    \end{split}
\end{equation}
which ties the prover's computation to the verifier's expectation.

Leveraging bilinearity, the equation simplifies as
\begin{equation}
    \begin{split}
        e(G, h)^{(H(\tau)Z(\tau) + C(\tau)) \cdot \rho_A \rho_B} 
        &= e(G, h)^{H(\tau)Z(\tau) \cdot \rho_A \rho_B} \\
        &\quad \cdot e(G, h)^{C(\tau) \cdot \rho_A \rho_B}
    \end{split}
\end{equation}
breaking down the computation and commitment into separable verifiable components. This culminates in the equality
\begin{align}
    e(\pi_H, vk_z) \cdot e(\pi_C, h) &= e(VK + \pi_A, \pi_B) \\
    \Rightarrow e(\pi_H, vk_z) \cdot e(\pi_C, h) &= e(VK + \pi_A, \pi_B)
\end{align}
confirming that the left and right sides of the pairing equation match, thus validating the proof.

In summary, the prover commits to a private flight path and generates a zk-SNARK proof attesting that they know a valid witness for the commitment $C_i$; the verifier, using the trusted setup's verification key and the proof components, confirms correctness through bilinear pairings, ensuring both privacy and integrity of the hidden path.

\subsection*{B. Security Analysis}

This section provides an in-depth security evaluation of the proposed protocol, covering essential security goals such as confidentiality, authentication, integrity, anonymity, non-repudiation, resilience against various attacks, and time synchronization.

\subsubsection*{1) 
 {Confidentiality}}

Confidentiality is achieved through ECDH for secure key exchange. Each participant, the user, the drone and the server, uses its own private key (e.g. \( r_1, r_2, r_3 \)) within an elliptical curve group to derive a common secret. This ensures that only legitimate entities can access the shared key, keeping the communication secure from eavesdroppers. Even if the communication is intercepted, the shared secret remains undiscoverable without access to the private keys. Additionally, zk-SNARKs protect sensitive details, such as the drone’s flight path, by allowing proof of compliance without revealing the actual path. This mechanism upholds flight and location privacy throughout the protocol's execution.

\subsubsection*{2) 
  {Authentication}}

Authentication is guaranteed through mutual identity verification among all entities involved, User, Drone, and Server. The protocol uses elliptic curve cryptography and digital signatures for this purpose. zk-SNARK-based proofs (e.g., \( \text{Auth}_u, \text{Auth}_d, \text{Auth}_s \)) confirm each party’s knowledge of secret credentials (e.g., \( I_u, I_d, I_s \)) without exposing those credentials. Moreover, hashed identifiers (e.g., \( H(PWD_u \parallel ID_u) \)) conceal user identities during communication, making it infeasible for attackers to deduce real identities from the messages.

\subsubsection*{3)  {Integrity}}

Integrity of data is preserved through robust cryptographic techniques. Every message exchanged (such as \( \text{Msg}_1, \text{Msg}_2, \text{Msg}_3 \)) contains hash values and authentication tags derived from earlier information, ensuring that any modification is detectable. zk-SNARKs (e.g., \( \pi_u, \pi_d, \pi_s \)) appended to the messages validate their authenticity and confirm that the sender has not been impersonated or the content tampered with. The Server also performs verification of these proofs, assuring that messages from both the Drone and User remain intact and authentic.

\subsubsection*{4) {Anonymity}}

Anonymity is maintained through a combination of hashed identifiers and zero-knowledge proofs. zk-SNARKs allow each participant to verify knowledge of sensitive data (such as identity or flight path) without exposing it, ensuring complete identity protection.  Hashing techniques like \( H(PWD_u \parallel ID_u) \) shield real identities during transmission. Furthermore, the protocol safeguards location privacy by hiding the drone’s actual travel path using zk-SNARK encryption. As a result, no party including potentially untrusted ones can infer where the drone is or has been during its operation.

\subsubsection*{5)  {Non-repudiation}}

Non-repudiation is enforced by ensuring that every action taken during the protocol is cryptographically provable. Each party retains evidence of their participation in the form of zk-SNARKs and authenticated message logs (e.g., \( \text{Msg}_5 \) through \( \text{Msg}_8 \)). These cryptographic artifacts serve as undeniable records, enabling any involved entity to demonstrate that a specific message was sent or received at a certain time, effectively preventing denial of involvement.

\subsubsection*{6) {Attack Resistance}}

The proposed protocol incorporates a variety of defense mechanisms to resist known attack vectors in secure communication environments.
\begin{itemize}

\item \text{Resistance to Man-in-the-Middle (MITM) Attacks:} The protocol employs ECDH for secure key agreement, where shared secrets (e.g., $SK_{u,s}$, $SK_{s,d}$) require both parties’ private keys. Intercepted public keys alone are insufficient for computation, making MITM decryption or forgery infeasible. Additionally, zk-SNARK proofs authenticate entities without exposing secrets, preventing impersonation or proof forgery.

\item {Replay Attack Mitigation:} Replay attacks are prevented through the use of secure timestamps \( T_1, T_2, T_3, T_4 \) embedded within each message. Each entity verifies that the received timestamp is within an acceptable time window, e.g., \( |T_2 - T_1| < \Delta T \). This temporal validation ensures that stale or previously valid messages cannot be resent to disrupt the protocol or gain unauthorized access.

\item {Resistance to impersonation attacks  :} The proposed protocol resists impersonation attacks using mutual authentication with tokens such as 
$\mathit{Auth}_u$, $\mathit{Auth}_{s,u}$, $\mathit{Auth}_{s,d}$, $\mathit{Auth}_d$, and $\mathit{Auth}_{d,u}$. 
These tokens are computed from ECDH-based session keys and bound to identities, nonces, and timestamps. 
Since only legitimate entities possess the private keys and session-specific secrets, attackers cannot generate valid tokens, effectively preventing impersonation.

\item \text{Resistance to Collusion Attack:} The protocol thwarts collusion by enforcing independently generated secrets for users ($V_u$), drones ($V_d$), and servers ($V_s$), along with ephemeral nonces ($r_1, r_2, r_3$). Since these values are never shared or reused, colluding parties cannot combine partial knowledge to reconstruct session keys (e.g., $SK_{u,s} = V_u \cdot V_s \cdot G$). zk-SNARKs ($\pi_u, \pi_d, \pi_s$) further protect privacy by verifying authentication and session setup without exposing sensitive inputs ($PWD_u$, $ID_d$, $w_s$). Thus, even a drone–server collusion cannot recover $V_u$ or forge user-specific proofs derived from $I_u = h(PWD_u \| ID_u \| SMK_u)$. This guarantees strong resistance against collusion during initialization.

\item {Resistance to Key Compromise Impersonation (KCI): }
In many authentication protocols, compromising one party's key allows an adversary to impersonate others to that party. In our protocol, KCI is mitigated by mutual authentication and the fact that proof generation in zk-SNARKs depends on knowledge of multiple secret values. Even with access to one entity’s private key, an attacker cannot forge valid proofs or compute session keys without the secrets from the other party.

\item {Message Tampering Detection: }
Each message includes a cryptographic hash of prior context and zk-SNARK proof elements (e.g., \( \pi_u, \pi_d, \pi_s \)) to ensure that any modification or injection of data is detectable. Tampering with even a single bit of a message will result in failed proof verification or hash mismatches, prompting the receiver to reject the message immediately.

\end{itemize}

\subsection*{C. Theoretical and Experimental Analysis} To evaluate the privacy protection and scalability of the proposed ZAPS protocol, we conducted a series of controlled experiments in a simulated UAV communication environment. These experiments examined the core privacy keys, resistance to trajectory reconstruction, session unlinkability, and potential zk-SNARK information leakage, together with scalability indicators, including CPU power consumption, CPU utilization, and communication overhead, as the UAV network size increased.

All experiments were performed on a Windows~11 machine equipped with an Intel Core~i7-12700H CPU, 16~GB RAM, and Python~3.11. The following Python libraries were used: \texttt{NumPy}, \texttt{Pandas}, \texttt{Matplotlib}, \texttt{Seaborn}, \texttt{scikit-learn}, and \texttt{dtaidistance}. 
Message exchange traces were synthetically generated to emulate UAV--GCS communications. For zk-SNARK proof simulation, pseudo-random proof byte arrays conditioned on the witness were used to enable leakage analysis without heavy cryptographic back-ends. Real proof integration can be performed using \texttt{Circom}/\texttt{snarkjs} or \texttt{ZoKrates} under the Window Subsystem for Linux (WSL).

Each simulated session consisted of an initialization phase (four fixed-size messages), per-waypoint proof transmissions, and a final termination message. Flight paths were generated with lengths in \(\{5, 8, 10, 15\}\) waypoints, and message sizes and inter-arrival times were randomized to reflect realistic wireless conditions.

\subsubsection*{1) {Aggregate Privacy Outcomes}} 
\begin{figure}[h]
    \centering
    \includegraphics[width=0.99\linewidth]{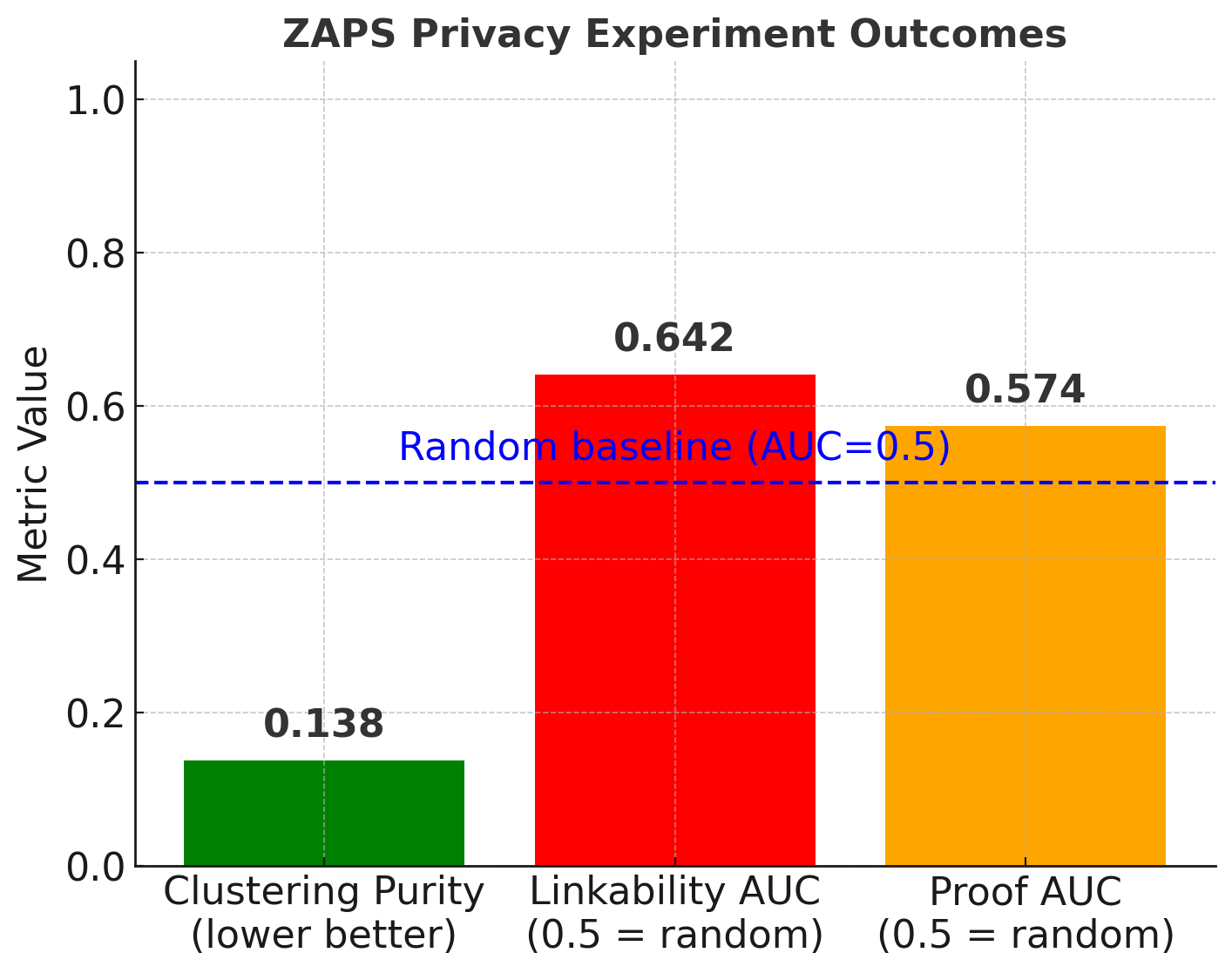}
   \caption {Summary of all privacy metrics (Purity for clustering, AUC for 
linkability and proof distinguishability) for the ZAPS protocol. 
\textit{Note:} For AUC-based attacks, $0.5$ indicates random guessing 
(no attack advantage). Values slightly above $0.5$ suggest limited leakage, 
whereas higher deviations from $0.5$ indicate stronger attack success.}
    \label{attack}
\end{figure}

{ The Aggregate Privacy Outcomes (Fig.~\ref{attack}) present a consolidated evaluation of ZAPS against three representative attack vectors: 
(1) \textit{Session Clustering Attack} -- grouping UAV communication sessions by metadata patterns such as timing and message sizes; 
(2) \textit{Linkability Attack} -- determining whether two different sessions originated from the same UAV; and 
(3) \textit{Proof Distinguishability Attack} -- identifying route-related details from zk-SNARK proof structures or byte distributions.}

To simulate these scenarios, synthetic UAV--GCS communication traces were generated with realistic message sizes, timing jitter, and route lengths. The adversary was modeled as a passive global observer with full access to all transmitted metadata but no plaintext content. Metrics used were clustering purity for the session clustering attack, classification accuracy for route inference, and AUC for linkability and proof distinguishability.

In the protected ZAPS configuration, session clustering purity was 0.138, only marginally above the random-mixing baseline ($\approx 0$). 
This shows that adversaries could group sessions only slightly better than chance.

The supervised session linkability test achieved an AUC of 0.642, which is only slightly above the 0.5 random baseline, indicating minimal correlation leakage. 
Similarly, proof distinguishability testing yielded an AUC of 0.574, again close to 0.5. Since $AUC = 0.5$ represents random guessing, values only slightly higher imply that adversaries gain no meaningful advantage from these attacks.

Together, these results confirm that ZAPS strongly resists session linking and proof-based inference, with only negligible statistical leakage.

\subsubsection*{ 2) {Scalability Analysis}} We simulated network scaling from 10 to 100 UAVs, measuring two key performance indicators: \textit{average handling time} (including proof verification and session processing) and \textit{communication size per UAV}. As shown in Fig.~6, both metrics exhibit near-linear growth with the number of UAVs, indicating predictable scaling behavior. Handling time increased from approximately 10.8~ms at 10 UAVs to 15.2~ms at 100 UAVs, remaining well within the operational requirements for real-time UAV coordination. Communication overhead grew from 118~KB to 220~KB per UAV over the same range, reflecting the proportional increase in per-UAV proof and message exchanges. This consistent scaling demonstrates that the protocol can accommodate substantial UAV swarm sizes without exceeding latency or bandwidth thresholds. Furthermore, the low absolute values of both metrics suggest that even deployments with hundreds of UAVs can be supported using commodity hardware and typical wireless data links, making the proposed ZAPS protocol viable for large-scale applications.

 \begin{figure}[h]
    \centering
    \includegraphics[width=0.99\linewidth]{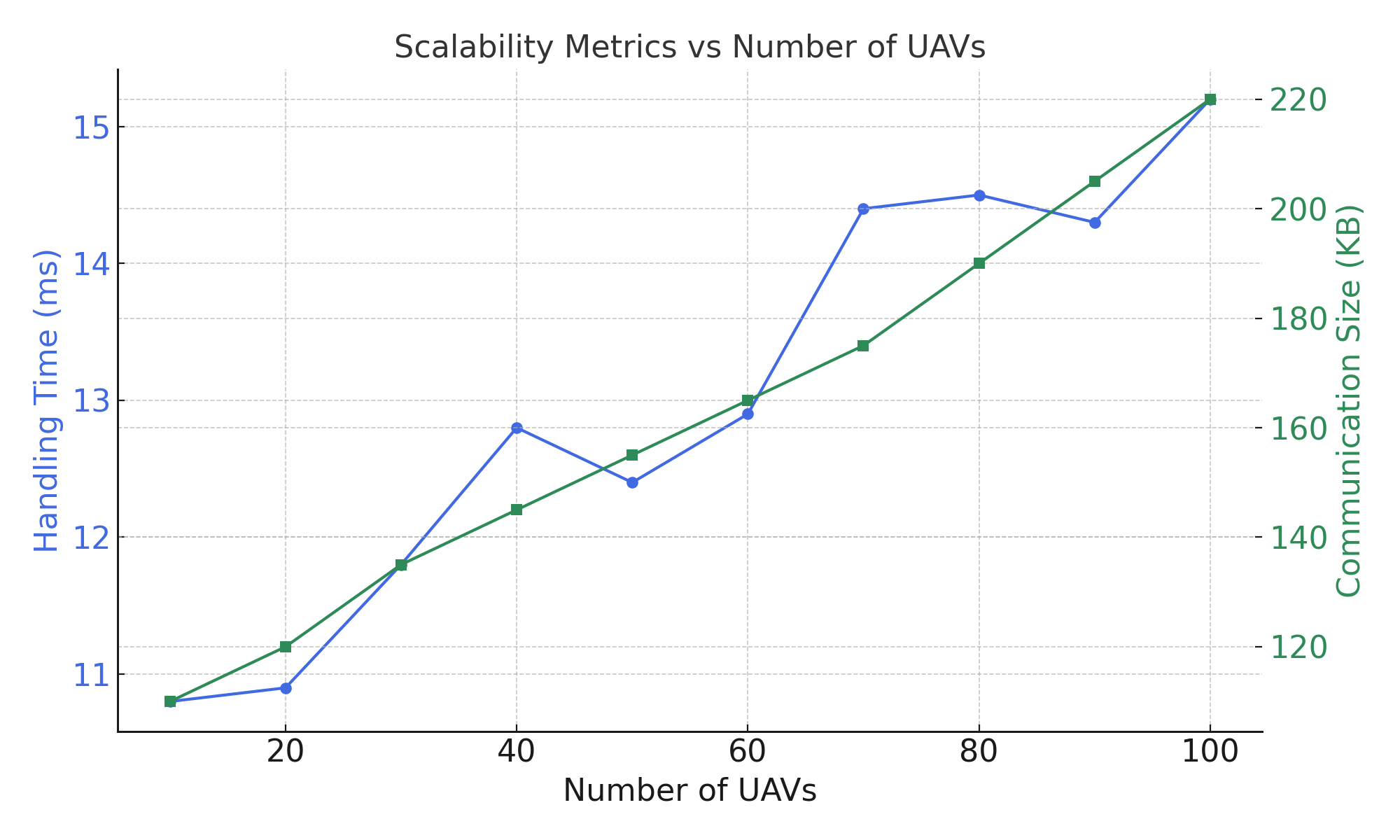}
    \caption{Scalability Metrics for Handling Time and Communication Size vs UAV count in the ZAPS protocol.}
    \label{sc}
\end{figure}

\subsubsection*{3)  {Communication Overhead }}  Table~\ref{communation overhead} shows that the communication overhead in the proposed drone delivery protocol is structured across two phases initialization and zk-SNARK Proof Generation accumulating to a total of 1,396 bytes per session. In the initialization phase (480 bytes), four messages are exchanged between the user, the drone, and the server to establish mutual authentication and the setup of sessions. These messages contain components such as elliptic curve public keys (32 bytes), identifiers (16 bytes), random nonces or hashes (32 bytes), and ECHD based authentication tokens (64 bytes), aligning with the assumptions of using secp256k1 for elliptic curve cryptography and SHA-256 for hashing. The zk-SNARK Proof Phase (916 bytes) transmits four additional messages, each embedding Groth16 zk-SNARK proofs (128 bytes), elliptic curve points, authentication tokens, and timestamps (4 bytes), enabling private yet verifiable proof of identity and flight path integrity. This carefully optimized structure balances strong cryptographic guarantees, such as anonymity, freshness, and zero-knowledge flight validation, against minimal communication overhead, making it suitable for privacy-sensitive and resource-constrained drone delivery applications.

\begin{table}[htbp]
  \centering
  \caption{Communication Overhead}
  \renewcommand{\arraystretch}{1.2} % slightly roomier rows
  \begin{tabular}{|c|l|c|}
    \hline
    \multicolumn{3}{|c|}{\textbf{1.\ Initialization Phase}} \\ \hline
    \textbf{Message} & \textbf{Components} & \textbf{Size (Bytes)} \\ \hline
    Msg$_1$ & $P_{u},\,RID_{u},\,I_{u},\,Auth_{u}$                   & $32 + 144 + 128$ \\ \hline
    Msg$_2$ & $Auth_{S},\,P_{S},\,I_{S},\,Auth_{S,u}$                & $16 + 162 + 62$  \\ \hline
    Msg$_3$ & $SID_{S},\,UID_{d},\,R_{S},\,Auth_{S,d}$               & $16 + 162 + 32$  \\ \hline
    \multicolumn{3}{|r|}{\textit{Initialization Total:}\; $80 + 144 + 128$} \\ \hline
    \multicolumn{3}{|c|}{\textbf{2.\ zk‑SNARK Proof Phase}} \\ \hline
    \textbf{Message} & \textbf{Components} & \textbf{Size (Bytes)} \\ \hline
    Msg$_5$ & $P_{u},\,RID_{u},\,I_{u},\,\pi_{u},\,Auth_{u}$         & $32 + 16 + 32 + 128$ \\ \hline
    Msg$_6$ & $Auth_{S},\,P_{S},\,I_{S},\,\pi_{S}$                   & $64 + 32 + 32 + 128$ \\ \hline
    Msg$_7$ & $P_{d},\,DID_{d},\,I_{d},\,\pi_{d},\,Auth_{d}$         & $32 + 16 + 32 + 128$ \\ \hline
    Msg$_8$ & $UID_{v},\,Auth_{u,S},\,I_{u,S}$                       & $16 + 64 + 32$       \\ \hline
    \multicolumn{3}{|r|}{\textit{zk‑SNARK Phase Total:}\; $276 + 256 + 272$} \\ \hline
    \multicolumn{3}{|c|}{\textbf{3.\ Total Communication Overhead}} \\ \hline
    \multicolumn{3}{|r|}{Total $= 480$ (Initialization) $+ 916$ (zk‑SNARK) $= 1{,}396$ bytes} \\ \hline
  \end{tabular}
    \label{communation overhead}
\end{table}

\subsubsection*{4)  {CPU Power Consumption Over Time}} 
\begin{figure}[h]
    \centering
    \includegraphics[width=0.99\linewidth]{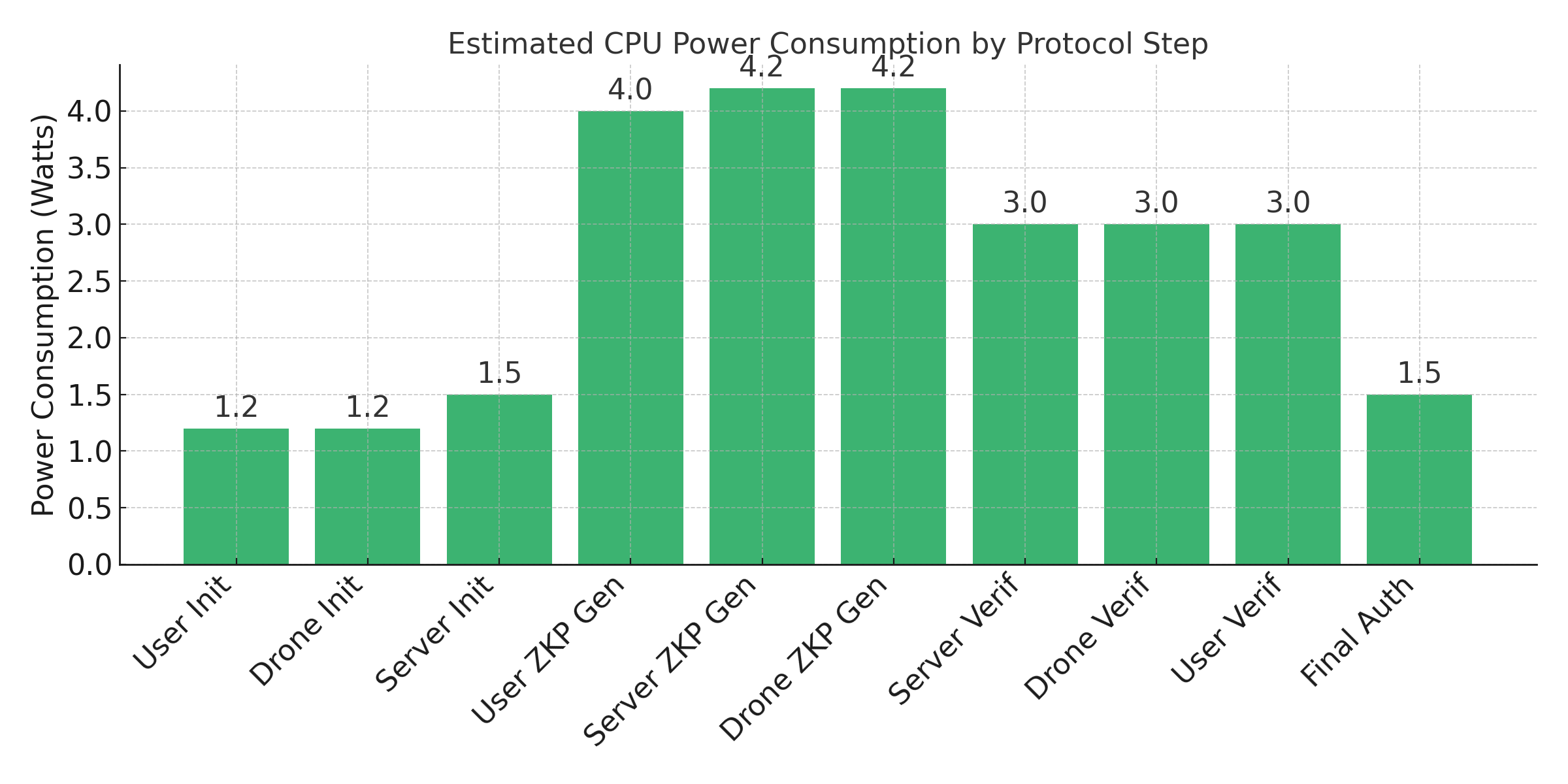}
    \caption{CPU power consumption over time for each step of ZAPS protocol.}
    \label{cpu power consumption}
\end{figure}

We adopt the profiles from \cite{ref39} as a basis for estimating CPU power consumption and CPU utilization over time.These works provide in-depth evaluations of CPU power consumption and processing demands in low-resource environments such as drones and embedded systems.The most power-intensive part of the protocol is the zk-SNARK proof generation phase performed by the User, Server, and Drone, which typically consumes between 4 and 4.2 watts due to the heavy cryptographic computations involved in generating zero-knowledge proofs. Verification steps executed by the Server, Drone, and User also demand considerable power, averaging around 3 watts. As shown in Fig.~\ref{cpu power consumption}. in contrast, other operations such as ECHD based key generation and final authentication are relatively lightweight, consuming only about 1.2 to 1.5 watts. When targeting constrained environments like drones, it may be beneficial to offload proof generation tasks or explore lightweight zk-SNARK variants to reduce power strain. From a comparative standpoint, the 3-watt verification steps represent approximately 5\% of the mean power consumption observed in typical performance graphs, while the lighter operations account for just 2--3\%. Overall, the power consumption profile of this zk-SNARK-based protocol is highly efficient, especially when contrasted with more resource demanding tasks, making it a viable option for privacy preserving drone applications and other energy sensitive platforms. 

\subsubsection*{3)  {CPU Utilization Over Time }}  
\begin{figure}[H]
    \centering
    \includegraphics[width=0.99\linewidth]{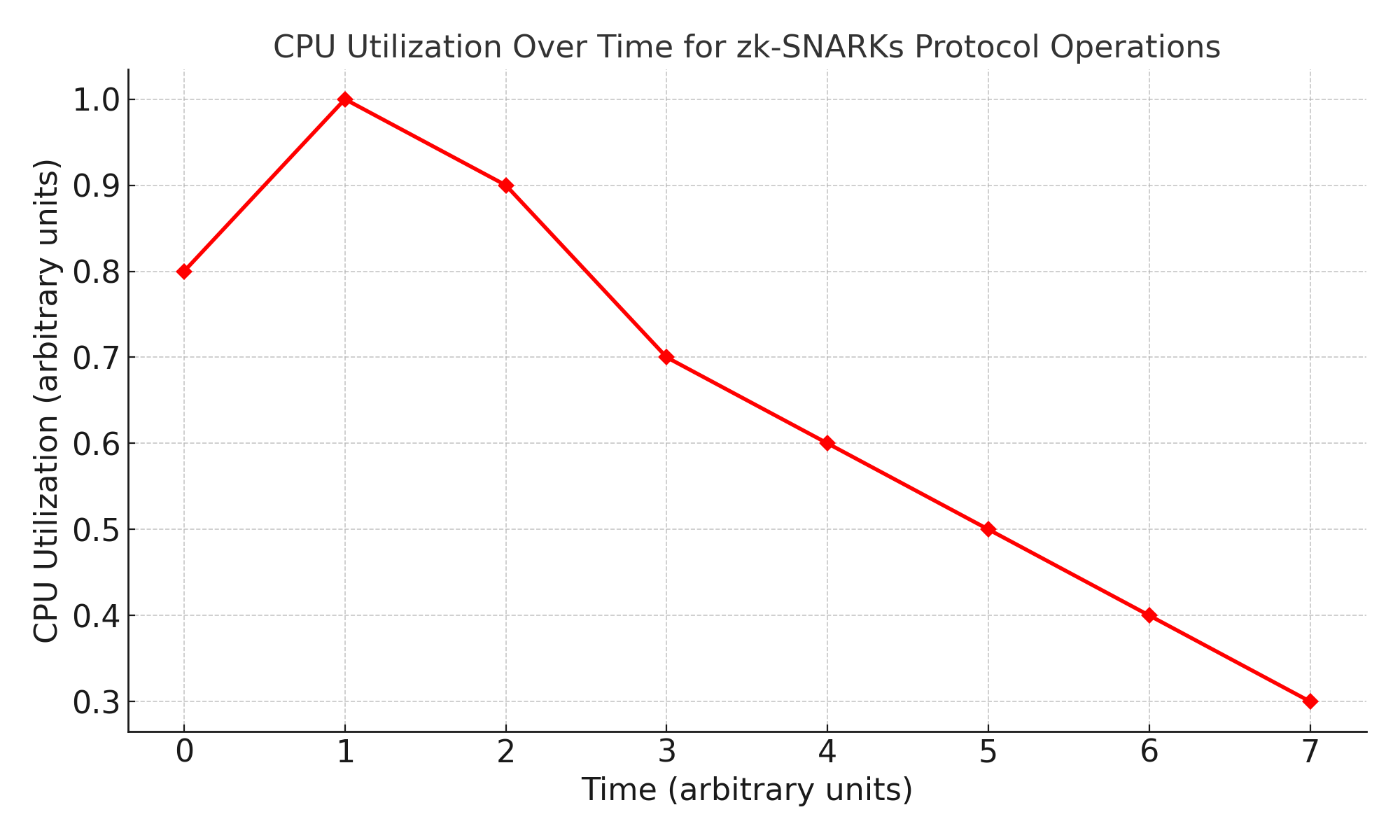}
    \caption{CPU utilization over time for ZAPS protocol.}
    \label{CPU Utilization}
\end{figure}

The Fig.~\ref{CPU Utilization}. illustrates CPU utilization Over Time for zk-SNARKs protocol operations, with the $x$ axis representing time in arbitrary units corresponding to different protocol phases and the $y$ axis showing relative CPU usage intensity. A red line with circular markers plots utilization values across time: starting at 0.8 (Time 0), peaking at 1.0 (Time 1), and then gradually declining to 0.3 by Time 7. This trend suggests that the initial phases, likely involving setup or proof generation, are computationally intensive, while subsequent steps such as verification are less demanding. The steady decline in CPU usage reflects a shift from heavy cryptographic operations to lightweight final steps like authentication. Overall, the plot effectively highlights how zk-SNARKs protocols impose high CPU demands early on, followed by a tapering load as the protocol progresses.

\section{Conclusion}
\change{This work presented ZAPS, a zero-knowledge proof-based authentication protocol for UAV systems designed to ensure flight path privacy while maintaining robust security against active and passive adversaries. By integrating zk-SNARKs into the UAV--GCS communication process, ZAPS enables mutual authentication and trajectory validation without disclosing sensitive flight path information. Theoretical analysis confirmed the protocol’s resilience against common threats, including replay, impersonation, and man-in-the-middle attacks. Experimental evaluation demonstrated strong privacy guarantees, with low clustering purity (0.138) for session unlinkability, minimal proof distinguishability (AUC 0.574), and only marginal linkability leakage (AUC 0.642). Scalability tests showed that ZAPS can support up to 100 UAVs with acceptable computational (15.2 ms average handling time) and communication ($<220$ KB per UAV) overhead. Overall, ZAPS delivers a privacy-preserving, scalable, and secure UAV authentication framework, making it a strong candidate for deployment in military, commercial, and civilian drone applications. Future work will focus on deploying ZAPS in real UAV hardware and extending its applicability to diverse domain-specific and cross-domain UAV operations, where location privacy, operational secrecy, and mission-specific confidentiality remain critical.}

% \section{REFERENCES}


\begin{thebibliography}{99}


\bibitem{ref1} M. Hassanalian and A. Abdelkefi, "Classifications applications and design challenges of drones: A review", \textit{Prog. Aerosp. Sci.}, vol. 91, pp. 99-131, May 2017.

\bibitem {ref2} W. Yang, S. Wang, X. Yin, X. Wang and J. Hu, "A Review on Security Issues and Solutions of the Internet of Drones," \textit{in IEEE Open Journal of the Computer Society}, vol. 3, pp. 96-110, 2022, doi: 10.1109/OJCS.2022.3183003.

\bibitem {ref3} M. Tanveer, H. Alasmary, N. Kumar, and A. Nayak, “SAAF-IoD: Secure
and anonymous authentication framework for the Internet of Drones,”
    \textit{IEEE Trans. Veh. Technol}., vol. 73, no. 1, pp. 232–244, Jan. 2024.
\bibitem{ref4} K. Mahmood, Z. Ghaffar, L. Nautiyal, M. Wahid Akram, A. Kumar Das and M. J. F. Alenazi, "A Privacy-Preserving Access Control Protocol for Consumer Flying Vehicles in Smart City Applications," \textit{IEEE Internet of Things Journal}, vol. 12, no. 1, pp. 978-985, 1 Jan.1, 2025. 

\bibitem{ref5} T. Zhaolu, Z. Wan and H. Wang, "Division of Regulatory Power: Collaborative Regulation for Privacy-Preserving Blockchains," \textit {IEEE Transactions on Information Forensics and Security}, vol. 19, pp. 2533-2548, 2024, doi: 10.1109/TIFS.2023.3348268.

\bibitem{ref6} K. Cho, M. Cho and J. Jeon, "Fly a Drone Safely: Evaluation of an Embodied Egocentric Drone Controller Interface,"  \textit {in Interacting with Computers}, vol. 29, no. 3, pp. 345-354, May 2017, doi: 10.1093/iwc/iww027.

\bibitem{ref7} C. Huang, Z. Ming and H. Huang, "Drone Stations-Aided Beyond-Battery-Lifetime Flight Planning for Parcel Delivery,"  \textit {in IEEE Transactions on Automation Science and Engineering}, vol. 20, no. 4, pp. 2294-2304, Oct. 2023, doi: 10.1109/TASE.2022.3213254.

\bibitem{ref8} Y. Miao, K. Hwang, D. Wu, Y. Hao and M. Chen, "Drone Swarm Path Planning for Mobile Edge Computing in Industrial Internet of Things," \textit {in IEEE Transactions on Industrial Informatics}, vol. 19, no. 5, pp. 6836-6848, May 2023, doi: 10.1109/TII.2022.3196392. 

\bibitem{ref9} T. Xia, M. Wang, J. He, L. Ni and G. Yang, "A AAV Swarm Authentication and Key Agreement Scheme Based on Latin Square Design," \textit {in IEEE Wireless Communications Letters}, vol. 14, no. 3, pp. 581-585, March 2025, doi: 10.1109/LWC.2024.3505903.

 \bibitem{ref10} B. Ma, X. Wang, X. Lin, Y. Jiang, C. Sun, Z. Wang, G. Yu, Y. He, W. Ni, R. P. Liu, " Location privacy threats and protections in future vehicular networks: A comprehensive review, " \textit {in arXiv preprint arXiv:2305.04503}, 2023/5/8.

\bibitem{ref11} U. Fiege, A. Fiat, and A. Shamir, “Zero knowledge proofs of identity,”
\textit { in Proceedings of the nineteenth annual ACM symposium on Theory of
 computing}, 1987, pp. 210–217.

\bibitem{ref12} Andola,N.;Raghav; Yadav, V.K.; Venkatesan, S.; Verma, S. SpyChain: A lightweight blockchain for authentication and anonymous
 authorization in IoD. \textit{ Wirel. Pers. Commun}. 2021, 119, 343–362.

 \bibitem{ref13} Pan, H.; Wang, Y.; Wang, W.; Cao, P.; Ye, F.; Wu, Q. Privacy-preserving location authentication for low-altitude UAVs: A
 blockchain-based approach. \textit{ Secur. Saf}. 2024, 3, 2024004.

\bibitem{ref14} Tai, W.-L.; Chang, Y.-F.; Li, W.-H. An IoT notion-based authentication and key agreement scheme ensuring user anonymity for
 heterogeneous ad hoc wireless sensor networks.  \textit{J. Inf. Secur. Appl}. 2017, 34, 133–141.
 \bibitem{ref15} Wazid, M.; Das, A.K.; Odelu, V.; Kumar, N.; Conti, M.; Jo, M. Design of Secure User Authenticated Key Management Protocol for
 Generic IoT Networks.  \textit{ IEEE Internet Things J}. 2018, 5, 269–282.

 \bibitem{ref16} Singh, J.; Gimekar, A.; Venkatesan, S. An efficient lightweight authentication scheme for human-centered industrial Internet of
 Things.  \textit{Int. J. Commun. Syst}. 2019, 2, e4189.
 
 \bibitem{ref17} Tian, Y.; Yuan, J.; Song, H. Efficient privacy-preserving authentication framework for edge-assisted Internet of Drones.  \textit{J. Inf. Secur. Appl}. 2019, 48, 102354.

\bibitem{ref18} Yue, X.; Liu, Y.; Wang, J.; Song, H.; Cao, H. Software defined radio and wireless acoustic networking for amateur drone
 surveillance. \textit{IEEE Commun. Mag}. 2018, 56, 90–97.

 \bibitem{ref19} Bouman,P.; Agatz, N.; Schmidt, M. Dynamic programming approaches for the traveling salesman problem with drone. \textit{Networks}
 2018, 72, 528–542.

 \bibitem{ref20} Shavarani, S.M.; Mosallaeipour, S.; Golabi, M.; ˙ Izbirak, G. A congested capacitated multi-level fuzzy facility location problem: An
 efficient drone delivery system. \textit{Comput. Oper. Res}. 2019, 108, 57–68.

\bibitem{ref21} Won,J.; Seo, S.-H.; Bertino, E. Bertino, Certificateless cryptographic protocols for efficient drone-based smart city applications.
\textit{ IEEE Int.J}. 2017, 5, 3721–3749.

\bibitem{ref22} Allouch, A.; Cheikhrouhou, O.; Koubâa, A.; Toumi, K.; Khalgui, M.; Nguyen, G.T. UTM-Chain: Blockchain-Based Secure
 Unmanned Traffic Management for Internet of Drones. \textit{Sensors} 2021, 21, 3049.

\bibitem{ref23}  Alsamhi, S.H.; Shvetsov A.V; Shvetsova S.V.; Hawbani A.; Guizani M.; Alhartomi M.A.; Ma O. Blockchain-Empowered Security
 and Energy Efficiency of Drone Swarm Consensus for Environment Exploration.  \textit{IEEE Trans. Green Commun. Netw}. 2023, 7,
 328–338. 
\bibitem{ref24} Koulianos, A.; Litke, A. Blockchain Technology for Secure Communication and Formation Control in Smart Drone Swarms.
\textit{ Future Internet} 2023, 15, 344.

\bibitem{ref25} Turkanovi´ c, M.; Brumen, B.; Hölbl, M. A novel user authentication and key agreement scheme for heterogeneous ad hoc wireless
 sensor networks, based on the Internet of Things notion.\textit{Ad Hoc Network} 2014, 20, 96–112.

\bibitem{ref26} Amin,R.; Islam, S.H.; Biswas, G.; Khan, M.K.; Leng, L.; Kumar, N. Design of an anonymity-preserving three-factor authenticated
 key exchange protocol for wireless sensor networks. \textit{Comput. Netw}. 2016, 101, 42–62.

\bibitem{ref27} Challa, S.; Wazid, M.; Das, A.K.; Kumar, N.; Reddy, A.G.; Yoon, E.-J.; Yoo, K.-Y. Secure signature-based authenticated key
 establishment scheme for future IoT applications. \textit{IEEE Int.J} 2017, 5, 3028–3043.

\bibitem{ref28} Gope,P.; Sikdar, B. An Efficient Privacy-Preserving Authenticated Key Agreement Scheme for Edge-Assisted Internet of Drones.
 \textit{IEEE Trans. Veh. Technol}. 2020, 69, 13621–13630. 
\bibitem{ref29} Zhang, Y.; He, D.; Li, L.; Chen, B. A lightweight authentication and key agreement scheme for Internet of Drones. \textit{Comput.
 Commun}. 2020, 154, 455–464.
 
 \bibitem{ref30} Ever, Y.K. A secure authentication scheme framework for mobile-sinks used in the internet of drones applications.\textit{ Comput.
 Commun}. 2020, 155, 143–149. 
 
\bibitem{ref31} Hussain, S.; Mahmood, K.; Khan, M.K.; Chen, C.M.; Alzahrani, B.A.; Chaudhry, S.A. Designing secure and lightweight user
 access to drone for smart city surveillance. \textit{Comput. Stand. Interfaces} 2021, 80, 103566.


 \bibitem{ref32} M. Hao, C. Shang, S. Wang, W. Jiang, and J. Nie, "UAV-Assisted Zero Knowledge Model Proof for Generative AI: A Multiagent Deep Reinforcement Learning Approach,"\textit{  IEEE Internet of Things Journal}, vol. 12, no. 10, pp. 13441–13454, 15 May 2025, doi: 10.1109/JIOT.2025.3531914.
 
\bibitem{ref33} M. Hao, C. Shang, S. Wang, W. Jiang and J. Nie, "UAV-Assisted Zero Knowledge Model Proof for Generative AI: A Multi-Agent Deep Reinforcement Learning Approach,"  \textit {IEEE Internet of Things Journal}, doi: 10.1109/JIOT.2025.3531914.

\bibitem{ref34}B.Parno, J. Howell, C. Gentry, and M. Raykova, Pinocchio: Nearly prac
tical veri able computation,\textit{ Commun. ACM}, vol. 59, no. 2, pp. 103112,
 2016.

\bibitem{ref35} J. M. Miret, D. Sadornil, and J. G. Tena, Pairing-based cryptography
 on elliptic curves, \textit{ Math. Comput. Sci.}, vol. 12, no. 3, pp. 309318,
 Sep. 2018.

 \bibitem{ref36} D. Dissanayaka, T. R. Wanasinghe, O. De Silva, A. Jayasiri and G. K. I. Mann, "Review of Navigation Methods for UAV-Based Parcel Delivery,"\textit{ in IEEE Transactions on Automation Science and Engineering}, vol. 21, no. 1, pp. 1068-1082, Jan. 2024, doi: 10.1109/TASE.2022.3232025

 \bibitem{ref37} Z. Deng, F. Wu, Y. Xu, D. Yang and L. Xiao, "Energy Minimization for Radio Map-Based UAV Pickup and Delivery Logistics System," \textit{ in IEEE Transactions on Vehicular Technology}, vol. 73, no. 11, pp. 17893-17898, Nov. 2024, doi: 10.1109/TVT.2024.3430320.

\bibitem{ref38} S. Naziri, X. Wang, G. Yu, J. Xu, S. Shrestha and C. J. Liang, "SMAKAP: Secure Mutual Authentication and Key Agreement Protocol for RFID Systems," \textit{ 2024 17th International Conference on Security of Information and Networks (SIN)}, Sydney, Australia, 2024, pp. 1-8, doi: 10.1109/SIN63213.2024.10871792.

\bibitem{ref39} A.Koulianos, P. Paraskevopoulos, A. Litke, Nikolaos K. Papadakis, Enhancing Unmanned Aerial Vehicle Security: A Zero-Knowledge Proof Approach with Zero-Knowledge Succinct Non-Interactive Arguments of Knowledge for Authentication and Location Proof, \textit{ Sensors} 2024, 24(17), 5838; https://doi.org/10.3390/s24175838

 
 

\end{thebibliography}
\end{document}